\documentclass{article}

\usepackage{amsfonts}
\usepackage{amsmath}
\usepackage{amssymb}
\usepackage{graphicx}
\usepackage{geometry}
\usepackage{hyperref}

\providecommand{\U}[1]{\protect\rule{.1in}{.1in}}

\begin{document}

\title{Exact Solutions for a Quantum Ring with a Dipolar Impurity}
\author{{Mourad Baazouzi}$^{1}$ \and {Mustafa Moumni}$^{1,2,a}$ \and {Mokhtar Falek}$^{1}$ \\
$^{1}$Laboratory of Photonic Physics and Nano-Materials (LPPNNM)\\
Department of Matter Sciences, University of Biskra, ALGERIA \\
$^{2}$Laboratory of Radiation and their Interactions with Matter (PRIMALAB)\\
Department of Physics, University of Batna1, ALGERIA\\
$^{a}$correspondant author m.moumni@univ-biskra.dz}
\maketitle

\begin{abstract}
We study analytically the Schr\"{o}dinger equation for a system made up of a quantum ring with a dipolar impurity under the effect an Aharonov-Bohm field and we choose the pseudoharmonic oscillator as a confinement potential. We calculate the exact values of the energies and we also get the exact expressions of the wave functions. We study the effects of the dipole moment of the impurity on the energies of the levels as well as on those of the transitions and this for different materials.
\end{abstract}

\tableofcontents

\section{Introduction}

The interest for two-dimensional $2D$ materials comes from the great popularity of the graphene (and co. like silicene and manganene) and also from experimental achievements with the realization of quantum gases at low dimensions \cite{Gorlitz01, Martiyanov10} and before that from quasi-condensate experiments \cite{Safonov98}. With the immense technological advancement in nano-processing, new beings appear in low
dimensional systems like quantum dots (QD) which can be regarded as low-dimensional hetero-structures whose carriers are confined in all spatial dimensions \cite{Chakraborty99}. Their manufacturing techniques make it possible to control their properties and thus they are made in such a way that they acquire the same characteristics of atomic systems; this is why they are sometimes called artificial atoms \cite{Chakraborty99, Pal19}. The confinement potential in QD may originate from various physical effects and possesses different symmetries in different nano-structures and the knowledge of realistic profile of confinement potential is necessary for a theoretical description of the electronic properties of QDs and, more importantly, for fabrication of nano-devices \cite{Pal19}. Since a long time, the pseudoharmonic oscillator (PHO) is used to study these compounds and recently it was found to be one of those that best correspond to QDs \cite{Liang19} (and the references therein).

For $2D$ disc-shaped quantum ring (QR) under the effect of an ionized donor atom, the conduction band electron is described by a PHO as a confinement potential and a donor impurity term \cite{Chen13, Niculescu17, Bejan18}:
\begin{equation}
V\left(r\right)=\frac{1}{2}m^{\ast}\omega_{0}^{2}r^{2}+\frac{\hbar^{2}}{2m^{\ast}}\frac{\lambda^{2}}{r^{2}}+\frac{e^{2}}{\varepsilon_{r}\left\vert
\overrightarrow{r}-\overrightarrow{r}_{j}\right\vert}
\label{eqt1}
\end{equation}

In the above expression, $m^{\ast}$ is the effective mass of the electron, $\varepsilon_{r}$ the static dielectric constant of the ring material, $\overrightarrow{r}$ is the position vector of the electron and $\overrightarrow{r}_{j}$ is the one of the impurity. $\lambda$ is a dimensionless parameter that characterizes the strength of the potential that describes the hollow region of the ring and $\omega_{0}$ represents the confinement frequency of the harmonic potential \cite{Bejan18, Bejan19}.

These Semiconductor (QR) can be made using many techniques. For instance, to get perfect circular QR, the most frequently used techniques are conventional molecular-beam epitaxy that creates a $2D$ electron gas in GaAs at the GaAs/GaAlAs interface, while asymmetric InGaAs QR can be made-up by solid-source beams \cite{Bejan19}.

The last term in \ref{eqt1} does not have a radial symmetry and so it adds a non-central term to the potential; this is the case for most real systems in quantum chemistry and nuclear physics like those used to describe ring-shaped molecules like benzene and also interactions between deformed pair of nuclei \cite{Zhang10}. The study of non-central potentials is not easy because analytically solvable cases are rare \cite{Hautot73, Berkdemir09, Moumni16, Kumari18, Parmar191, Parmar192} and this has led studies to use radial approximations to find analytical solutions or to consider numerical methods for computations.

In this work, we are interested in the analytical resolution of the $2D$ Schr\"{o}dinger equation (SE) for a QR with a new non-central PHO:
\begin{equation}
V(r,\theta)=Ar^{2}+\frac{B}{r^{2}}+C+D\frac{\cos\theta}{r^{2}}
\label{eqt2}
\end{equation}

We use this general form for the potential because the first three terms represent the PHO with the following substitutions $A=D_{e}r_{e}^{-2}$, $B=D_{e}r_{e}^{2}$ and $C=-2D_{e}$ where $D_{e}$ represents the dissociation energy and $r_{e}$ is the equilibrium internuclear separation \cite{Ikhdair08, Oyewumi12, Dong07}:
\begin{equation}
V_{PHO}(r,\theta)=D_{e}\left(\frac{r}{r_{e}}-\frac{r_{e}}{r}\right)^{2}
\label{eqt3}
\end{equation}
and also because it gives the first two terms of \ref{eqt1} when $A=m^{\ast}\omega_{0}^{2}/2$, $B=\hbar^{2}\lambda^{2}/2m^{\ast}$ and $C=0$. Another advantage of the general form in \ref{eqt2} is that it corresponds to the Tan-Inkson model of QR if we put $C=-2\sqrt{AB}$ \cite{Tan96}.

The last term in \ref{eqt2} is the dipole potential which comes from the multipolar expansion of the donor term in \ref{eqt1} \cite{Moumni11}. If we consider that the donor impurity is not ionized, the dipole is the first term in this expansion since there is no Coulomb (monopolar) contribution in this case. Our choice here adds a new case to that of the usual Coulomb impurities \cite{Pal19, Chen13, Lee13, Feng17, Yuan18, Naghdi18, Yaseen19, Tedondje20}.

In $2D$ systems, the dipole potential is present in ultrathin semiconductor layers \cite{Zhou90}, in spin-polarized atomic hydrogen absorbed on the surface of superfluid helium \cite{Vasilyev02}, in the states of planar charged particles with perpendicular magnetic field \cite{Gadella11} and in gapped graphene with two charged impurities \cite{Martino14, Klopfer14}. It corresponds also to the interaction with a permanent electric dipole moment in CdSe QD \cite{Cristea16} and to the electron pairing that stems from the spin-orbit interaction in $2D$ quantum well \cite{Gindikin18}.

We add to the PHO potential \ref{eqt2} a vector potential of the Aharonov-Bohm (AB) type as a lot of previous works \cite{Santander11, Bueno14, Oliveira19, Ghosh20}, but we will keep here the analytical character of the solutions.

Our work is structured as follows: after this first section which represents the introduction, there will be a second section \ref{sec:2DS} dedicated to the analytical solutions of the SE for the new non-central PHO potential \ref{eqt2}. A third section \ref{sec:QR} will follow where we apply our results to QRs and finally we will conclude our work in the forth section \ref{sec:CC}.

\section{Exact Solutions for 2D Schr\"{o}dinger Equation}
\label{sec:2DS}

We are looking for the solutions of the stationary $2D$ SE when a charged particle $q$ is subject to the effects of the non-central scalar potential $V\left(r,\theta\right)$ defined by \ref{eqt2} and a vector potential of Aharonov-Bohm (AB) $\overrightarrow{A}_{AB}$:
\begin{equation}
\left[\frac{1}{2\mu}\left(i\hbar\overrightarrow{\nabla}+e\overrightarrow{A}_{AB}\right)^{2}+qV\left(r,\theta\right)\right]\psi\left(r,\theta\right) =E\psi\left(r,\theta\right)\text{   ;   }\overrightarrow{A}_{AB}=\frac{\phi_{AB}}{2\pi r}\overrightarrow{e}_{\theta}
\label{eqt4}
\end{equation}

Since the AB field satisfies the Coulomb gauge $\overrightarrow{\nabla}\cdot\overrightarrow{A}_{AB}=0$, we have:
\begin{equation}
\left(i\hbar\overrightarrow{\nabla}+e\overrightarrow{A}_{AB}\right)^{2}\psi\left(r,\theta\right)=\left(-\hbar^{2}\Delta+e^{2}A_{AB}^{2} +2ie\hbar\overrightarrow{A}_{AB}\cdot\overrightarrow{\nabla}\right)\psi\left(r,\theta\right)
\label{eqt5}
\end{equation}

Because of the shape of the potential \ref{eqt2}, It is more convenient to use the polar coordinates ($r$,$\theta $) for Laplace operator and we write the SE of our system as:
\begin{equation}
\left[-\frac{\hbar^{2}}{2\mu}\Delta+\frac{e^{2}\phi_{AB}^{2}}{8\pi^{2}\mu r^{2}}+i\frac{e\hbar\phi_{AB}}{2\mu\pi r^{2}}\frac{\partial}{\partial\theta} +\left(Ar^{2}+\frac{B}{r^{2}} +\frac{D_{\theta}\cos\theta}{r^{2}}\right)\right]\psi\left(r,\theta\right)=(E-C)\psi\left( r,\theta\right)
\label{eqt6}
\end{equation}
With the use of $\varepsilon=2\mu\hbar^{-2}(E-C)$ and $\beta=B+\left(\hbar^{2}/2\mu\right)\left(\phi_{AB}^{2}/\phi_{0}^{2}\right)$ ($\phi_{0}=h/e$ is the magnetic flux quanta) we have:
\begin{equation}
\left[\frac{\partial^{2}}{\partial r^{2}}+\frac{1}{r}\frac{\partial}{\partial r}-\frac{2\mu A}{\hbar^{2}}r^{2}-\frac{2\mu\beta}{\hbar^{2}r^{2}} +\frac{1}{r^{2}}\left(\frac{\partial ^{2}}{\partial\theta^{2}}-2i\frac{\phi_{AB}}{\phi_{0}}\frac{\partial}{\partial\theta}-\frac{2\mu D_{\theta}}{\hbar^{2}}\cos\theta\right)\right]\psi\left(r,\theta\right)=\varepsilon\psi \left(r,\theta\right)
\label{eqt7}
\end{equation}

We use the following convention for the wave function $\psi(r,\theta)=r^{-1/2}R(r)\Theta(\theta)$ to split this equation into two parts:
\begin{subequations}
\begin{gather}
\left(\frac{d^{2}}{d\theta^{2}}-2i\frac{\phi_{AB}}{\phi_{0}}\frac{d}{d\theta}-E_{\theta}-\frac{2\mu D_{\theta}}{\hbar^{2}} \cos\theta\right)\Theta(\theta)=0
\label{eqt8a} \\
\left[\frac{d^{2}}{dr^{2}}+\left(E_{\theta}-\frac{2\mu\beta}{\hbar^{2}}+\frac{1}{4}\right)\frac{1}{r^{2}}-\frac{2\mu A}{\hbar^{2}}r^{2}+\varepsilon\right]R(r)=0
\label{eqt8b}
\end{gather}

We have to find the angular eigenvalues $E_{\theta }$ from the angular equation and then use these expressions to resolve the radial part and here we get the $\varepsilon$; this give us the energies $E$ of the system from $\varepsilon=2\mu\hbar^{-2}(E-C)$ and also the eigenfunctions $\psi(r,\theta)$ from $\psi(r,\theta)=r^{-1/2}R(r)\Theta(\theta)$.

\subsection{Solution of the Angular Equation}

For the angular equation, we write the solutions $\Theta\left(\theta\right)=\exp\left(i\delta\theta\right)\Phi\left(\theta\right)$ where $\delta=\phi_{AB}/\phi_{0}$ to obtain a Mathieu type equation \cite{Mathieu68} with the transformations $\theta=2z$, $c=4\left(\delta^{2}-E_{\theta}\right)$ and $p=4\mu\hbar^{-2}D_{\theta}$:
\end{subequations}
\begin{equation}
\frac{\partial^{2}\Phi(z)}{\partial z^{2}}+\left[c-2p\cos(2z)\right]\Phi(z)=0
\label{eqt9}
\end{equation}

The phase factor $\exp\left(i\delta\theta\right)$ is similar to that appearing when studying the electronic states near the Fermi level of a nanotube with sufficiently large diameter \cite{Ajiki93}.

The period of $z$ is $\pi$ because $\theta$ is $2\pi$ periodic, so the solutions of \ref{eqt9} are the cosine-elliptic $ce_{2m}$ and the sine-elliptic $se_{2m+2}$, where $m$ is a natural number \cite{Jazar20}. For a given value of the parameter $p$ (or equivalently $D_{\theta}$), the solutions are periodic only for certain values of the characteristic values $c$, denoted $a$ for $ce_{2m}$ solutions while those related to $se_{2m+2}$ functions are denoted $b$ \cite{Floquet83, Bloch28}.

For example in the case of small values of $p$ and for $m>3$, we have ($\nu=4m^{2}-1$) \cite{Jazar20}:
\begin{equation}
a_{2m}\approx b_{2m}=4m^{2}+\frac{1}{2\nu}p^{2}+\frac{20m^{2}+7}{32\nu^{3}\left(\nu-3\right)}p^{4}+\frac{36m^{4} +232m^{2}+29}{64\nu^{5}\left(\nu -3\right)\left(\nu-8\right)}p^{6}+O\left(p^{8}\right)
\label{eqt10}
\end{equation}

We find similar approximative polynomial forms for large $p$'s but with different coefficients \cite{Jazar20}.

We see that at the limit $D\rightarrow 0$ (or $p\rightarrow 0$), the two values are equal to $4m^{2}$ and so the solutions $ce$ and $se$ are degenerate. Since these solutions give the $\cos $ and $\sin $ functions respectively at this limit, we find the usual $\exp\left(im\theta\right)$ solution found for central potentials. In our case, and because of the non-central term $r^{-2}\cos\theta$ in the potential, this degeneracy is removed and a new phase factor is added $\exp\left(i\delta\theta\right)$ to the one coming from the Floquet theorem $\exp\left(im\theta\right)$. Through this phase factor, the Mathieu values $m$ are shifted to $m^{\prime}=m+\delta$ \cite{Santander11, Oliveira19, Ferkous13, Neto18}.

Using the relations of $c_{2m}$ and $p$, we get the angular eigenvalues $E_{\theta}$ as a function of $D_{\theta}$:
\begin{equation}
E_{\theta}^{(m,\delta)}=\delta^{2}-\frac{1}{4}c_{2\left(m+\delta\right)}\left(\frac{4\mu}{\hbar^{2}}D_{\theta}\right)\approx\left(\frac{\phi
_{AB}}{\phi _{0}}\right)^{2}-\left(m+\frac{\phi_{AB}}{\phi_{0}}\right)^{2}+P_{\left(m+\phi_{AB}/\phi_{0}\right)}(D_{\theta})
\label{eqt11}
\end{equation}

The last relation is an approximate expression of $E_{\theta}$ for small values of $D_{\theta}$ where $P(D_{\theta})$ is a polynomial of even powers of $D_{\theta }$ starting from the power $2$; It connects to the quantum number $m$.

We use the same symbol $c_{2m}$ in the $E_{\theta }$ expression \ref{eqt11} to denote both $a_{2m}$\ and $b_{2m}$ to simplify the notations and from now on, this will be the case in all the relations. Because $c_{2m}\left(p\rightarrow 0\right)=4m^{2}$\ref{eqt10}, these Mathieu characteristic values generalise the quantum number $m$ in the same way as the $ce$ and $se$ solutions generalize the trigonometric functions $\cos$ and $\sin$. But $ce$ and $se$ are not degenerate like $\cos$ and $\sin$ since $a_{2m}\neq b_{2m}$ except for $D=0$.

It is interesting to note that the same angular equation \ref{eqt8a} appears when we consider the motion of charged particles on a $1D$ circular ring in a double-well potential and an AB flux \cite{Kettemann99}. We find also a similar $\phi_{AB}/\phi_{0}$ phase factor in the case of an AB ring described with Dirac equation \cite{Oliveira19, Neto18}.

\subsection{Solution of the Radial Equation}

Now having in mind the values of $E_{\theta }^{(m,\delta )}$, we start the resolution of the radial equation \ref{eqt8b}. Do do it, we put $r=a\sqrt{\rho}$, $a^{2}=\hbar/\sqrt{2\mu A}$ and $\eta=\left(E_{\theta}-2\mu\beta\hbar^{-2}+1/4\right)$, so we get:
\begin{equation}
\left[4\rho\frac{d^{2}}{d\rho^{2}}+2\frac{d}{d\rho}-\rho+\frac{\eta}{\rho}+\varepsilon a^{2}\right]R(\rho )=0
\label{eqt12}
\end{equation}

To solve this equation, we use the following transformation:
\begin{equation}
R(\rho)=\rho^{\alpha}e^{-\rho/2}\omega\left(\rho\right)
\label{eqt13}
\end{equation}
So we get a new differential equation for $\omega(\rho)$:
\begin{equation}
\left[\rho\frac{d^{2}}{d\rho^{2}}+\left(2\left(\alpha+\frac{1}{4}\right)-\rho\right)\frac{d}{d\rho}-\rho +\frac{1}{\rho}\left(\left(\alpha-\frac{1}{4}\right)^{2}-\frac{1-4\eta}{4}\right)+\frac{\varepsilon a^{2}}{4}-\left(\alpha+\frac{1}{4}\right)\right]\omega=0
\label{eqt14}
\end{equation}

We use that the fact that $\alpha$ is a free parameter to cancel the term in $1/\rho$, so we put:
\begin{equation*}
\left(\alpha-\frac{1}{4}\right)^{2}-\frac{1-4\eta}{4}=0
\end{equation*}

Since we require $\omega\left(\rho\right)$ to be nonsingular at $\rho\rightarrow 0$, we choose the positive solution for $\alpha$:
\begin{equation}
\alpha=\frac{1}{2}\left(\frac{1}{2}+\sqrt{1-4\eta}\right)
\label{eqt15}
\end{equation}

Finally we put:
\begin{equation}
4n_{r}=\varepsilon a^{2}-4\alpha -1
\label{eqt16}
\end{equation}
to simplify the equation \ref{eqt14} to the form:
\begin{equation}
\left(\rho\frac{d^{2}}{d\rho^{2}}+\left(2\alpha+\frac{1}{2}-\rho\right)\frac{d}{d\rho}+n_{r}\right)\omega\left(\rho\right)=0
\label{eqt17}
\end{equation}
The solutions of this equation are the hypergeometric functions ${}_{1}F_{1}$ ($N$ is the normalized constant):
\begin{equation}
\omega\left(\rho\right)=N{}_{1}F_{1}\left(-n_{r},2\alpha+1/2,\rho\right)\text{ ; }n_{r}\in\mathbb{N}
\label{eqt18}
\end{equation}

In terms of the variables $r$ and $\theta$, the general form of the wave function $\psi(r,\theta)$ is:
\begin{equation}
\psi(r,\theta)=N\left(\frac{r}{a}\right)^{2\alpha}e^{-r^{2}/2a^{2}}{}_{1}F_{1}\left(\left(\alpha+\frac{1}{4}\right)-\frac{\varepsilon a^{2}}{4},2\alpha+\frac{1}{2},\frac{r^{2}}{a^{2}}\right)
\label{eqt19}
\end{equation}

To determine the normalization constant $N$, we substitute this expression into the condition $\int\left\vert\psi\left(r,\theta\right) \right\vert^{2}rdrd\theta=1$ and we recall that $\Theta(\theta )$ is normalized to $\pi $ by definition \cite{Jazar20}. We use Laguerre polynomials of degree $n$ from the relation \cite{Abra72}:
\begin{equation}
L_{n}^{\left(2\alpha-1/2\right)}\left(\frac{r^{2}}{a^{2}}\right)=\frac{\left( n+2\alpha-1/2\right)!}{n!\left(2\alpha-1/2\right)!} {}_{1}F_{1}\left(-n,2\alpha+\frac{1}{2},\frac{r^{2}}{a^{2}}\right)
\label{eqt20}
\end{equation}
and the identity \cite{Grad07}:
\begin{align}
\int\limits_{0}^{\infty}q^{k+1/2}e^{-q}L_{n}^{k}\left(q\right)L_{m}^{k}\left(q\right)dq=&\frac{\Gamma\left(n+k+1\right)^{2}\Gamma
\left(m+k+1\right)\Gamma\left(k+3/2\right)\Gamma\left(m-1/2\right)}{n!m!\Gamma\left(k+1\right)\Gamma\left(-1/2\right)}\times  \notag \\
& {}_{3}F_{2}\left(-n,k+3/2,3/2;k+1,-m+3/2;1\right)
\label{eqt21}
\end{align}
to get the formula:
\begin{align}
N&=\frac{\left(n+2\alpha-1/2\right)!}{\left(2\alpha-1/2\right)!a\Gamma\left(n+2\alpha+1/2\right)}\times  \notag \\
&\left[\frac{2\Gamma\left(2\alpha+1/2\right)\Gamma\left(-1/2\right)}{\Gamma\left(n+2\alpha+1/2\right)\Gamma\left(2\alpha+1\right)
\Gamma\left(n-1/2\right)_{3}F_{2}\left(-n,2\alpha+1,3/2;2\alpha+1/2,-n+3/2;1\right)}\right]^{1/2}
\label{eqt22}
\end{align}
One can also use the general result of the PHO in $N$ dimensions \cite{Oyewumi12} by setting $N=2$ and replacing $m$ by $\sqrt{c_{2m}}/2$ \ref{eqt10} and thus obtain a more compact version of this constant.

For the energies, we use the equations \ref{eqt15} and \ref{eqt16} and the relations $\eta=E_{\theta}-2\mu\beta\hbar^{-2}+1/4$, $\beta=B+\left(\hbar^{2}/2\mu\right)\delta^{2}$, $E_{\theta}=\delta^{2}-(1/4)c_{2\left(m+\delta\right)}$ and $\varepsilon=2\mu\hbar^{-2}(E-C)$ to get the eigenvalues:
\begin{equation}
E=\sqrt{\frac{2\hbar^{2}A}{\mu}}\left(2n_{r}+1+\sqrt{\frac{2\mu\beta}{\hbar^{2}}-E_{\theta}}\right)+C =\sqrt{\frac{2\hbar^{2}A}{\mu}}\left(2n_{r}+1+\sqrt{\frac{c_{2\left(m+\delta\right)}}{4}+\frac{2\mu B}{\hbar^{2}}}\right)+C
\label{eqt23}
\end{equation}

By setting $\delta=0$ here, we find the energies for the new non-central dipolar PHO \ref{eqt2}:
\begin{equation}
E=\sqrt{\frac{2\hbar^{2}A}{\mu}}\left(2n_{r}+1+\sqrt{\frac{c_{2m}}{4}+\frac{2\mu B}{\hbar^{2}}}\right)+C
\label{eqt24}
\end{equation}

and by setting $D=0$, we obtain the case of the Tan-Inkson model \cite{Tan96} with an AB field:
\begin{equation}
E=\sqrt{\frac{2\hbar^{2}A}{\mu}}\left(2n_{r}+1+\sqrt{\left(m+\frac{\phi_{AB}}{\phi_{0}}\right)^{2}+\frac{2\mu B}{\hbar^{2}}}\right)+C
\label{eqt25}
\end{equation}

We can also write the energies \ref{eqt23} according to the parameters $D_{e}$ and $r_{e}$:
\begin{equation}
E_{n_{r},m}=\sqrt{\frac{2\hbar ^{2}D_{e}}{\mu r_{e}^{2}}}\left( 2n_{r}+1+\sqrt{\frac{2\mu}{\hbar^{2}}D_{e}r_{e}^{2} +\frac{1}{4}c_{2\left(m+\delta\right)}\left(\frac{4\mu}{\hbar^{2}}D_{\theta}\right)}\right)-2D_{e}
\label{eqt26}
\end{equation}

These general relations give the usual PHO energies \cite{Oyewumi12} for $D_{\theta }=0$ because $c_{2m}=4m^{2}$  in this case from \ref{eqt10} ($\delta =0$ as there is no AB field) and also the usual harmonic oscillator solutions if we add the assumptions $A=\mu\omega^{2}/2$ and $B=C=0$ to a null dipole moment:
\begin{equation}
E=\hbar\omega\left(2n_{r}+\left\vert m\right\vert+1\right)
\label{eqt27}
\end{equation}

\section{Applications to 2D Quantum Rings}
\label{sec:QR}

To apply these results to the two-dimensional QR, we use the notations of \ref{eqt1}, so we have $A=m^{\ast}\omega_{0}^{2}/2$, $B=\lambda^{2}\hbar^{2}/2m^{\ast}$ ($m^{\ast}=\mu$) and $C=0$ in \ref{eqt23} to write the energies as follows:

\begin{equation}
E_{n_{r},m}=\hbar\omega_{0}\left(2n_{r}+1+\sqrt{\lambda^{2}+\frac{1}{4} c_{2\left(m+\delta\right)}\left(\frac{4m^{\ast}}{\varepsilon_{r}\hbar^{2}}D\right)}\right)
\label{eqt28}
\end{equation}
We have also written $D_{\theta}=D/\varepsilon_{r}$ to introduce the same parameters of the QR potential \ref{eqt1}.

Focusing on the effects of the dipole moment on the energies, we see in their expression \ref{eqt28} that the main modification is due to the parameter $c_{2m}$ which replaces $m$; So we will study the effects on these energies through the root term:

\begin{equation}
\lambda_{ce,se}=\sqrt{\lambda^{2}+\frac{1}{4}c_{2\left(m+\delta\right)}\left(\frac{4m^{\ast}}{\varepsilon_{r}\hbar^{2}}D\right)}
\label{eqt29}
\end{equation}
and especially through the correction that $D$ adds to this value and therefore we will focus on the dimensionless difference $\lambda_{ce,se}-\lambda_{0}$ where $\lambda _{0}=\sqrt{\lambda ^{2}+m^{2}}$ since il will give us also the corrections of the energies in $\hbar \omega _{0}$ units. The indices $ce$ and $se$ in \ref{eqt29} indicate that the corrections depend on the chosen solution type for the angular equation \ref{eqt8a}.

Parameter values used in our computations correspond to GaAs devices where $\lambda =2$, $m^{\ast}=0.067m_{e}$, $\varepsilon_{r}=12.65$ \cite{Bejan18, Bejan19, Paspalakis13, Bejan17, Jayarubi19} and we use the Hartree atomic units defined by $\hbar=e=m_{e}=4\pi\varepsilon_{0}=1$. For the energy numerical values, we have $\hbar\omega_{0}\approx 0.1\sim1eV$ \cite{Bejan18, Bejan19, Bejan17, Evangelou19} and this means that the energies of the levels considered in our work ($n=1\text{\&}2$ and $m=0\text{\&}1$) are in the intervals $0.5$ to $0.8eV$ or $5$ to $8eV$ depending on the value of $\hbar\omega_{0}$. For $D$, we choose them in the range $1$ to $10 a.u.$ because it corresponds to the experimental values of most molecular systems \cite{Nelson67}.

\begin{figure}[tbp]
\centering
\includegraphics[width=0.5\textwidth]{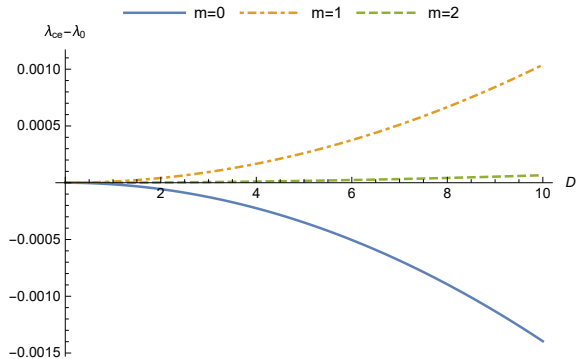}
\caption{Corrections of $ce$ energies in $\hbar \protect\omega_{0}$ units
for $m=0,1,2$}
\label{Fig1}
\end{figure}

\begin{figure}[tbp]
\centering
\includegraphics[width=0.5\textwidth]{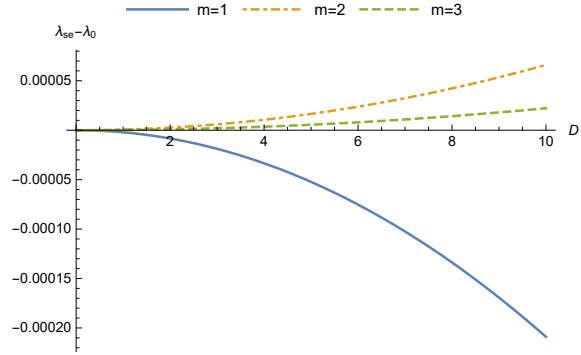}
\caption{Corrections of $se$ energies in $\hbar \protect\omega_{0}$ units
for $m=1,2,3$}
\label{Fig2}
\end{figure}

Because of the behavior of the Mathieu characteristic values $a_{2m}$ and $b_{2m}$, the corrections for the $ce$ states $m=0$ and the $se$ states $m=1$ are negative, while they are positive for all the other states for both $ce$ and $se$ solutions ($m=0$ states exist only for $ce$ solutions). Their values decrease with increasing $m$ and those corresponding to $ce$ solutions are larger than those of $se$ ones for the same quantum numbers (Figs \ref{Fig1} and \ref{Fig2}). These figures show that we can neglect the modifications for $m\geq 2$ as they are $10^{2}$ smaller than those corresponding to the $s$-states ($m=0$) and so they give corrections of the order of $10^{-5}eV$ or less. Depending on the values of $\hbar\omega_{0}$ mentioned above, the energy corrections for $m=0$ are around $10^{-3}eV$ while those corresponding to $m=1$ are just a little bit smaller for $ce$ states and approximatively equal to $10^{-4}eV$ for $se$ states.

\begin{figure}[tbp]
\centering
\includegraphics[width=0.5\textwidth]{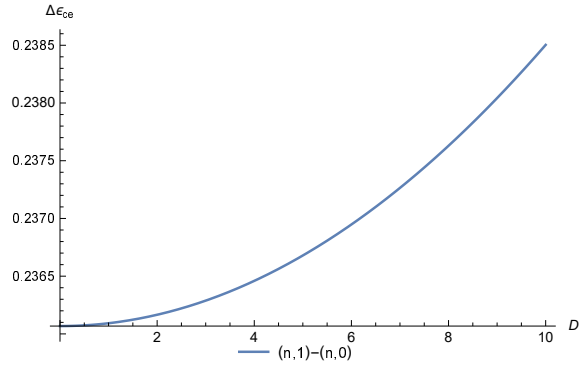}
\caption{Correction of the transitions $(n,1)\rightarrow(n,0)$ in $\hbar
\protect\omega_{0}$ units for $ce$ solutions}
\label{Fig3}
\end{figure}

\begin{figure}[tbp]
\centering
\includegraphics[width=0.5\textwidth]{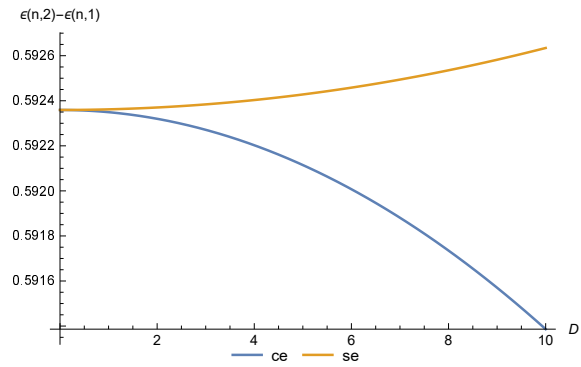}
\caption{Correction of the transitions $(n,2)\rightarrow(n,1)$ in $\hbar
\protect\omega_{0}$ units for $ce$ and $se$ solutions}
\label{Fig4}
\end{figure}

Since these corrections are not the same for the different values of $m$, the dipolar term modifies the transition energies between the levels; in Figs \ref{Fig3} and \ref{Fig4}, we give as an example the effects on the transitions $(n,1)\rightarrow(n,0)$ and $(n,2)\rightarrow (n,1)$. Note that the presence of the dipole term increases the transition energies by more than $1\%$ in the case of $(n,1)\rightarrow(n,0)$ while it decreases that of $(n,2)\rightarrow (n,1)$ by about $0.1\%$; this concerns $ce$ states. Regarding the $se$ states, its presence increases the energy of the $(n,2)\rightarrow (n,1)$ transition by less than $0.04\%$.

\begin{figure}[tbp]
\centering
\includegraphics[width=0.5\textwidth]{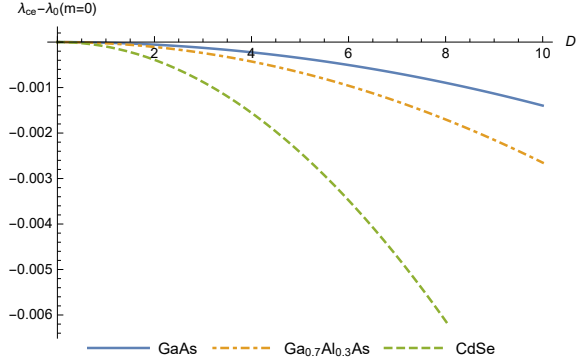}
\caption{Corrections for some materials of $ce$ energies in $\hbar \protect\omega_{0}$ units for $m=0$}
\label{Fig5}
\end{figure}

\begin{figure}[tbp]
\centering
\includegraphics[width=0.5\textwidth]{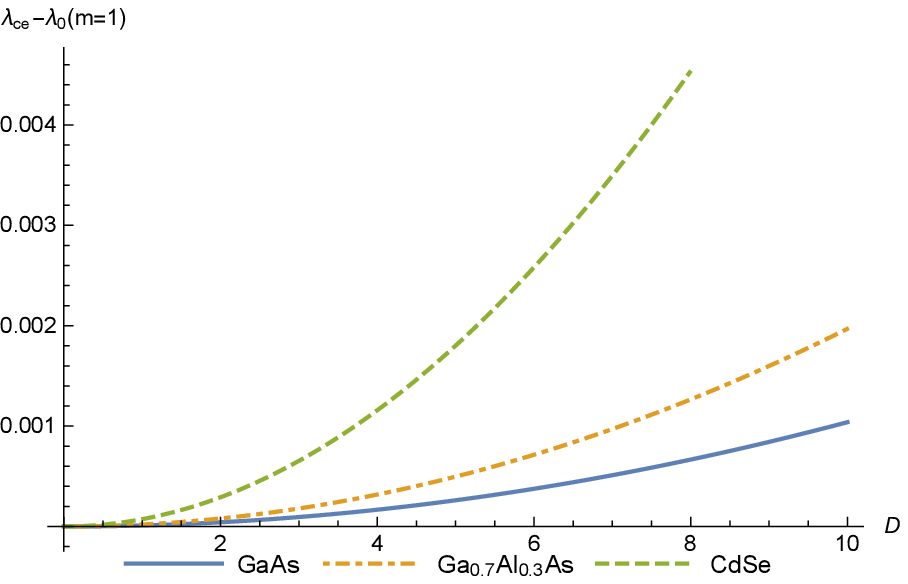}
\caption{Corrections for some materials of $ce$ energies in $\hbar \protect\omega_{0}$ units for $m=1$}
\label{Fig6}
\end{figure}

\begin{figure}[tbp]
\centering
\includegraphics[width=0.5\textwidth]{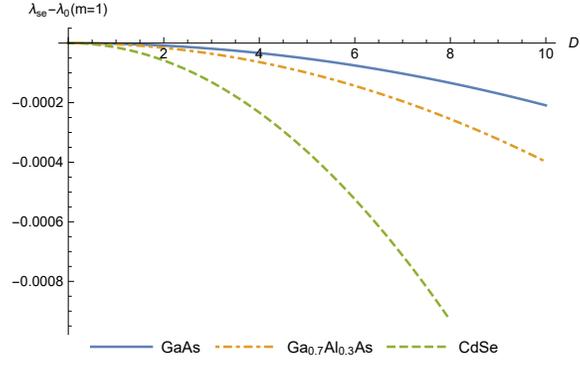}
\caption{Corrections for some materials of $se$ energies in $\hbar \protect\omega_{0}$ units for $m=1$}
\label{Fig7}
\end{figure}

From \ref{eqt29}, we see also that the corrections increase with the ratio $m^{\ast}/\varepsilon_{r}$ and thus they are more pronounced for the compounds $Ga_{1-x}Al_{x}As$, since the effective mass for these materials is given by the formula $m^{\ast}=(0.067+0.085x)m_{e}$\ with $x$ real \cite{Chrafih19}. We show in Figs \ref{Fig5}, \ref{Fig6} and \ref{Fig7}, these changes for $x=0.3$ used in \cite{Bejan17, Chrafih19} and also for the parameters of $CdSe$ $m^{\ast}/\varepsilon_{r}=0.13/9.3$ studied in \cite{Cristea16}. We observe that the dipole corrections are $2$ times greater for the $Ga_{1-x}Al_{x}As$ than for the $GaAs$ and they are $7$ times more pronounced than the latter in the case of $CdSe$.

\begin{figure}[tbp]
\centering
\includegraphics[width=0.5\textwidth]{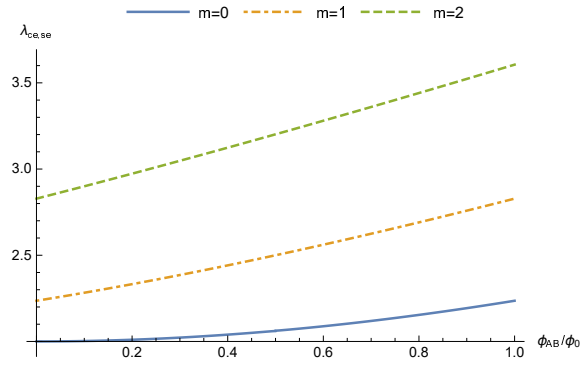}
\caption{Corrections of $ce$ and $se$ energies in $\hbar\protect\omega_{0}$ units from AB field}
\label{Fig8}
\end{figure}

Concerning the effects of the AB field, they are very large compared to those due to the dipole potential since they appear directly in the Mathieu exponent while the dipole moment appears in the Mathieu parameter (This is why we choose $\Delta=0$ in the precedent figures \ref{Fig1} to \ref{Fig7} representing the effects of $D$ on the energies); the AB corrections are of the order of $10^{-1}\hbar\omega$. We notice that they increase the energies of the system and they are equal for both solutions "ce" and "se". These corrections also increase with increasing $m$ (Figs \ref{Fig8}).

\section{Conclusion}

\label{sec:CC}

In this work, we have analytically solved the Schr\"{o}dinger equation for a quantum ring confined by a pseudoharmonic potential and under the effect of a dipolar impurity and an Aharonov-Bohm field. We have obtained the exact expressions of the energies for this new non-central potential as well as those of the wave functions by using the Mathieu functions $ce$ and $se$ in the angular part of the solutions. Our calculations show that the field AB is present through a phase in the wave functions and so it shifts the quantum number $m$ of the system. These corrections increase the energies and they are much larger than those due to the dipole.

The first characteristic of the dipole term is that it removes the degeneracy present for central potentials; thus the energies depend on the orientation of the solutions compared to the dipole direction, which broke the central symmetry by becoming a privileged one. Corrections are more pronounced for "ce" states and therefore states whose orientations are in the same direction as the dipole; this is similar to the dependence of $3D$ energies on the azimuth number $m$ as soon as we are in the presence of a Hamiltonian term depending on the direction like a constant magnetic field.

Our solutions generalize the azimuthal quantum number $m$ through the Mathieu characteristic values. The corrections are larger for $m=0$ and they decrease as it increases; this generates a correction on the transition energies between the different levels and it is more apparent for those between the lowest ones as $(n,1)\rightarrow (n,0)$ and $(n,2)\rightarrow (n,1)$. All these corrections depends on the chosen material and we show that they are proportional to the ratio of the effective mass on the static dielectric constant $m^{\ast}/\varepsilon_{r}$.

Finally we mention that if we add another vector potential $Br/2$ to the AB one, we will no longer have the possibility of treating the system analytically and we will then have to use perturbative methods, which is beyond the scope of this work as we are looking for analytical solutions.

\section*{Acknowledgment}

This work was done with funding from the DGRSDT of the Ministry of Higher Education and Scientific Research in Algeria as part of the PRFU
B00L02UN070120190003.

\end{document}